\documentclass[12pt]{article}%

\usepackage{amsmath}
\usepackage{graphicx}
\usepackage{amsfonts}
\usepackage{amssymb}

\begin{document}

\title{\bf What do very nearly flat detectable cosmic topologies look like?}
\author{
B.Mota$^1\,$\thanks{brunom@cbpf.br} ,
G.I. Gomero$^2\,$\thanks{german@ift.unesp.br} ,
M.J. Rebou\c{c}as$^1\,$\thanks{reboucas@cbpf.br} , \ and \
R. Tavakol$^3\,$\thanks{r.tavakol@qmul.ac.uk} \\
\\
$^1$ Centro Brasileiro de Pesquisas F\'\i sicas, \\
Rua Dr. Xavier Sigaud 150 \\
Rio de Janeiro, RJ 22290--180, Brazil \\
\\
$^2$ Instituto de F\'{\i}sica Te\'orica, Universidade Estadual Paulista, \\
Rua Pamplona 145 \\
S\~ao Paulo, SP 01405--900, Brazil \\
\\
$^3$ Astronomy Unit, School of Mathematical Sciences, \\
Queen Mary, University of London, \\
Mile End Road, London E1 4NS, UK
}

%\date{\today}

\maketitle

\begin{abstract}
\noindent
Recent studies of the detectability of cosmic topology of nearly flat
universes have often concentrated on the range of values of $\Omega_{0}$ given
by current observations. Here we study the consequences of taking the bounds on
$\Omega_{0}$ given by inflationary models, i.e. $|\Omega_0 - 1| \ll 1$. We show
that in this limit, a generic detectable non-flat manifold is locally indistinguishable
from either a cylindrical ($\mathbb{R}^2 \times \mathbb{S}^1$) or toroidal
($\mathbb{R} \times \mathbb{T}^2$) manifold, irrespective of its global shape,
with the former being more likely. Importantly this is compatible with some
recent indications based on the analysis of high resolution CMB data. It also
implies that in this limit an observer would not be able to distinguish
topologically whether the universe is spherical, hyperbolic or flat.
By severely restricting the expected topological signatures of detectable
isometries, our results provide an effective theoretical framework for
interpreting cosmological observations, and can be used to confine any
parameter space which realistic search strategies, such as the `circles in the
sky' method, need to concentrate on. This is particularly important in the
inflationary limit, where the precise nature of cosmic topology becomes
undecidable.

\end{abstract}

Two fundamental questions regarding the nature of the universe concern its
geometry and topology. Regarding geometry, recent high precision data by WMAP%
~\cite{WMAP} have provided strong evidence suggesting that the universe is
nearly flat, with the ratio of its total matter--energy density to the critical
value, $\Omega_{0}$, very close to one. They, however, do not fix the sign of
its curvature.\footnote{The most recent estimates of the density parameters
specify this region to be $\Omega_{0} \in[0.99,1.05]$ and $\Omega_{\Lambda}
\in[0.69,0.79]$ with a $2\sigma$ confidence, which still leaves open the sign
of the curvature of the cosmic geometry.} The near flatness of the universe is
compatible with inflationary models which further predict this ratio to be
extremely close to one. Regarding topology, most recent studies of the
detectability of the cosmic topology have so far concentrated on the range of
values of $\Omega_{0}$ given by current observations~\cite{grt,weeks}. These
works have demonstrated that the likelihood of detectability of the cosmic
topology depends crucially on the curvature radius.

Here we take the bounds predicted by inflationary models, where $|\Omega_{0}
- 1|$ is orders of magnitude smaller than those indicated by current
observations, and examine what would a detectable topology look like in such
nearly flat universes. We study the local nature of detectable hyperbolic and
spherical manifolds and discuss the significance of our results for search
strategies for the detection of the topology of the universe.

We assume that the universe can be locally modelled by a
Friedmann-Lema\^{\i}tre-Robertson-Walker (FLRW) metric and consider the
possibility that the $3$-space may be a multiply connected manifold, $M =
\widetilde{M} /\Gamma$, where $\widetilde{M} = \mathbb{S}^3$ or $\mathbb{H}^3$,
and $\Gamma$ is a fixed-point free and discrete group of isometries of
$\widetilde{M}$.

Since we study the topology of universes with curved spatial sections, we need
to express the redshift--distance relation in units of the curvature radius,
which depends on the matter--energy content of the universe as well as
$\Omega_0$. Even though current observations favor the content to be well
approximated by dust plus a dark energy, its actual details do not affect our
considerations below, and thus in this sense our results are robust. This is
important since the precise nature of the dark energy is at present not known.

A natural way to study the detectability of topology is through the lengths of
closed geodesics. For an isometry $g\in\Gamma$ and a point $x\in M$, the length
of the closed geodesic generated by $g$ is given by its distance function
$d(x,gx)$, i.e. the distance between $x$ and its image $gx$. This readily
allows the definition of the injectivity radius $r_{inj}(x)$ as half the length
of the smallest closed geodesic passing through $x$. One can also define an
injectivity radius for the whole manifold $M$ as $r_{inj} = \inf_{x \in M}
r_{inj}(x)$, which is the radius of the smallest sphere inscribable in $M$. A
necessary condition for detectability of cosmic topology is then given by
$r_{inj}(x) < \chi _{obs}$, where $\chi_{obs}$ is the redshift--distance
relation evaluated at the maximum redshift ($z=z_{max}$) of the survey used.
Similarly, a sufficient condition for detectability is $r_{max} < \chi_{obs}$,
where $r_{max} = \sup_{x \in M}r_{inj}(x)$ is the maximum injectivity radius.

To study the local shape of the detectable non--flat manifolds in the density
limit suggested by inflationary models, $|\Omega_0 - 1| \ll 1$, we make three
physically motivated assumptions: (i) the observer is at a position $x$ where
topology is detectable, (ii) the survey depth is very small in units of the
curvature radius, and (iii) the topology is not excludable, i.e. it does not
produce too many images so as to make it already observationally excludable.%
\footnote{
A sufficient condition for a topology to be excludable is $r_{max} \ll
\chi_{obs}$.
}
Therefore in the following our main physical assumption will be
\begin{equation}
r_{inj}(x) \lesssim \chi_{obs} \ll 1 \;. \label{hip}%
\end{equation}

An important class of isometries are the Clifford translations (CT), defined
as those with constant distance functions. We shall encounter isometries whose
distance functions do not vary appreciably inside the observable sphere of
radius $\chi_{obs}$. This motivates the definition of an isometry $g$ as
CT--like inside a given (here the observable) sphere if the difference between
the maximum and minimum lengths of all closed geodesics generated by $g$ that
lie within the sphere is sufficiently small ($\ll1$). In the following we
consider the hyperbolic and spherical spaces separately, and prove that in the
limit (\ref{hip}) all isometries that generate observable images generically
behave CT--like.

It is known that the set of compact orientable hyperbolic manifolds can
be ordered in a sequence of sequences of manifolds. The manifolds in
each individual sequence are ordered by increasing volume, with a
cusped manifold as its limit. This implies that for a sufficiently
large index $i$ a compact manifold $M_{i}$ looks like a cusped manifold,
and we shall therefore refer to such manifolds as cusped--like. An
important feature of such cusped and cusped-like manifolds is that they
possess regions (namely the cusp regions) where $r_{inj}(x)$ takes small
values (zero in the limiting case). Therefore for a given $\chi_{obs}$,
in any sequence we can always choose $i$ such that for all $j>i$, the
cusped--like regions of the manifolds $M_{j}$ satisfy (\ref{hip}). Thus
the compact hyperbolic manifolds that have detectable topologies are
the cusped-like ones in their cusp regions.

Such cusped--like regions are well approximated by pure cusp manifolds, whose
covering group is generated by two parabolic isometries with the same fixed
point at infinity. Since these isometries commute, the fundamental group of a
pure cusp is $\mathbb{Z} \times \mathbb{Z}$,  thus, topologically, a pure cusp
is equivalent to $\mathbb{R} \times \mathbb{T}^2$. In principle, the length
scales of both generators are mutually independent, however if one of the
length scales is larger than the diameter of the observable sphere, then the
detectable part of the topology will be that of a horn, which is equivalent to
$\mathbb{R}^2 \times \mathbb{S}^1$. Within all known cusped--like manifolds,
pure cusps generically have one length scale much smaller than the other, which
therefore makes the latter case more likely.

We can formulate the previous argument more precisely by studying the CT--like
behavior of a parabolic isometry in the limit (\ref{hip}). A convenient model
of a hyperbolic space for our analysis is the upper--half space model,
$\mathbb{H}^3 = \{x=(x_1,x_2,x_3) \in \mathbb{R}^3\; : \; x_3>0\}$. The
distance between two points $x,y \in \mathbb{H}^3$ in this model is given by
\begin{equation}
\cosh(d(x,y)) = 1 + \frac{|x-y|^2}{2x_3y_3} \; , \label{hypdist}%
\end{equation}
and a parabolic isometry can always be put in the form $gx = (x_1+L, x_2,
x_3)$, with $L>0$, so that, under the condition (\ref{hip}), its distance
function reduces to $d(x,gx) = \frac{L}{x_3}$.

We can obtain a bound on the variation of this distance function inside a
detectable sphere of radius $\chi_{obs}\ll1$ by computing the difference
between the maximum and minimal lengths of closed geodesics associated with
$g$ inside this sphere. Up to second order this gives
\begin{equation}
\frac{\Delta d}{d_0} < \frac{1}{2} \, \frac{d_{\max} - d_{\min}}{d_0} \simeq
\chi_{obs} \; ,
\end{equation}
where $d_0$ is the distance function of $g$ evaluated at the center of the
sphere, and $\Delta d = |d(x,g(x))-d_0|$. This demonstrates that under the
condition (\ref{hip}), parabolic isometries behave CT--like inside the
observable universe, thus an observer living in a horn or cusped region, may
not distinguish, even topologically, her universe as being hyperbolic or flat.

We shall now consider the spherical spaces and recall that any spherical
3-manifold is a quotient $\mathbb{S}^3/\Gamma$, where $\Gamma$ is a f\/inite
subgroup of $SO(4)$ acting freely on the $3$-sphere. The classif\/ication of
spherical $3$-dimensional manifolds is well known~\cite{Sphere}. Since $\Gamma$
is a finite group, any element $g \in\Gamma$ is of finite order, thus $\Gamma$
contains cyclic subgroups. Generically, the injectivity radius is determined by
the cyclic subgroup of largest order of $\Gamma$~\cite{Weeks}. We shall
therefore initially consider the action of cyclic groups on $\mathbb{S}^3$.

A cyclic group $\mathbb{Z}_p$ may act on $\mathbb{S}^3$ in different ways
parametrized by an integer $q$ such that $p$ and $q$ are relatively prime, and
$1\leq q < p/2$. These actions give rise to the lens spaces $L(p,q)$, whose
global injectivity radii depend only on $p$ and are
given by $r_{inj} = \frac{\pi}{p}$. The lens spaces $L(p,1)$ are globally
homogeneous, thus in the following we shall only consider inhomogeneous lens
spaces ($q \ge 2$ and $p \ge 5$).

Representing the 3-sphere as the subset of pairs $z=(z_1,z_2)$ of complex
numbers of unit length, $|z|^2 = |z_1|^2 + |z_2|^2 = 1$, let $g_{(p,q)}$ be
the generator of the covering group of $L(p,q)$, and $g^nz$ the image of $z$
by its $n$-th iteration. The distance between $z$ and its $n$-th image
is then given by
\begin{equation}
\cos(d(z,g^nz)) = \cos \! \left(\frac{2\pi n}{p}\right) |z_1|^2 + \cos \!
\left(\frac{2\pi nq}{p}\right) |z_2|^2\;.\label{cos}%
\end{equation}

In order to determine the local nature of a lens space we shall require some
results from number theory (see e.g.~\cite{NT}).%
\footnote{
Any rational number $\frac{q}{p}<1$, with $q$ and $p$ relative primes, can be
written as a continued fraction
\[
\frac{q}{p} = \frac{1}{a_{2}+\frac{1}{a_{3}+F_4}} \; ,
\]
with $F_n < 1$, thus allowing the representation $\frac{q}{p} \equiv
[0,a_2,a_3,...,a_k]$ with $a_{i} \in \mathbb{N}$, which is unique if
we demand $a_{k}>1$. The convergents are defined as $c_i =
\frac{q_i}{p_i} = [0,a_2,a_3,...,a_i]$ for $i \leq k$, where $q_i$
and $p_i$ which are relative primes can be obtained recursively, the
sequences $\{q_i\}$ and $\{p_i\}$ are strictly increasing and $p_{k-1}
< \frac{p}{2}$. It can further be shown that
\begin{equation}
\frac{q}{p}-\frac{q_i}{p_i} = \frac{(-1)^{i+1}}{p_i (p_{i+1} + p_i
F_{i+2})} \equiv (-1)^{i+1} \frac{G_i}{p_i} \; . \label{conv}
\end{equation}
}
We start by computing the iterate of $g_{(p,q)}$ which generates the smallest
geodesic at a given point $z$. For points on the great circle $(z_1,0)$ the
smallest geodesic is the one corresponding to the generator (or its inverse).
This, however, is not the case for points on the great circle $(0,z_2)$. In
fact, in this case the $n$-th iteration of $g_{(p,q)}$ will generate the
smallest geodesic if $n$ is chosen such that $nq=\pm1\!\mod\!(p)$. The
existence of this integer is guaranteed by Euler's theorem, according to which
$q^{\phi(p)} = 1 \! \mod \! (p)$, where $\phi(p)$ is the number of integers
which are relative prime to and smaller than $p$. Thus, the smallest geodesic
for any point on $(0,z_2)$ will be generated by the choice of $n = \pm
q^{\phi(p)-1}\mod\!(p)$. Moreover, the only points for which $r_{inj}(z) =
\pi/p$ lie on the great circles given by $(z_1,0)$ and $(0,z_2)$. These circles
are the equators of minimum injectivity radius, and will be important in the
following considerations.

It can be shown that the following generalization holds: For any point in the
lens space $L(p,q)$ there is a convergent $\frac{q_j}{p_j}$ such that the
isometry that generates the smallest geodesic is $g^{p_j}$. An immediate
consequence of this is that the number of candidate isometries that may
generate the smallest geodesic is $k$. Furthermore, $k \leq \log_24q^2$ and
thus $k$ is small even for lens spaces with large $p$ and $q$. Therefore, this
result provides an effective procedure for obtaining the length of the smallest
geodesic at any point. Moreover, using Eq.~(\ref{conv}) from the footnote, and
recalling that for any integer $m$, $\cos(2\pi m+s) = \cos(|s|)$, we have
\begin{equation}
\cos \left( 2\pi\frac{p_i q}{p} \right) = \cos \left( \frac{2\pi}{p}
\left \vert p_i q - pq_i \right \vert \right) = \cos(2\pi G_i) \; ,
\label{dd}
\end{equation}
with $i=1,...,k$. Thus, since $G_k = 1/p$, for points on the equator $(0,z_2)$
the isometry which generates the smallest geodesics is $g^{p_{k-1}}$, and
therefore $q^{\phi(p)-1} \equiv \pm p_{k-1} \operatorname{mod}(p)$.
Now For any $n=p_j$, we can rewrite (\ref{cos}) as
\begin{equation}
\cos(d(z,g^{p_j}z)) = \cos\left( \frac{2\pi p_j}{p} \right) |z_1|^2 +
\cos(2\pi G_j) |z_2|^2 \, . \label{15}
\end{equation}
Choosing $j$ from now on such that $p_j \leq \sqrt{p} \, \leq \, p_{j+1}$ we
have $\frac{2\pi p_j}{p} \leq \frac{2\pi}{\sqrt{p}}$ and $2\pi G_j \leq
\frac{2\pi}{\sqrt{p}}$. To obtain an upper bound on the length of the smallest
closed geodesics at any point, we substitute these inequalities into (\ref{15})
obtaining $\cos(d(z,g^{p_j}z)) \geq \cos \left( \frac{2\pi}{\sqrt{p}} \right)$,
for all $z$ in the 3-sphere. We then have
\begin{equation}
r_{\max}\leq \frac{1}{2} \sup_{z\in M}d(z,g^{p_j}z) \leq \frac{\pi}{\sqrt{p}}
= r_{inj} \sqrt{p} \label{bound}
\end{equation}
since $r_{\max}$ is the supremum of the smallest geodesics. A similar bound can
be obtained if we restrict our analysis to geodesics generated by $g_{(p,q)}$
(see \cite{grt}). The minimum length of such geodesics is $\leq\frac{2\pi
q}{p}$, and this can be viewed as an alternative upper bound on $2r_{\max }$.
It is clear that $r_{\max} \leq \frac{\pi q}{p}$ is a better bound than
(\ref{bound}) if and only if $q<\sqrt{p}$. Moreover, in this case, since $p_2
= \operatorname{int}[\frac{p}{q}]$, we have $p_2 > \sqrt{p}$, thus $p_j = p_1
= 1$ and therefore the generator $g_{(p,q)}$ gives the smallest geodesic
at least for $|z_2|\le|z_1|$, which corresponds to half of the lens space.

In the above analysis no constraints were imposed on the lens spaces, i.e.
the values of $p$ and $q$.
We wish now to concentrate on values of these parameters which lead to observable
isometries in the limit (\ref{hip}). In order to show that the
isometry that generates the smallest geodesic behaves CT--like, it suffices to
show that $g^{p_j}$ behaves CT--like in the observable universe. Recall that
any detectable lens space will have $p > \frac{\pi}{\chi_{obs}}$ \cite{grt},
thus from (\ref{bound}) we have that $d(z,g^{p_j}z)$ is small. Expanding
(\ref{15}) and using the fact that $|z_1|^2 + |z_2|^2 = 1$ we obtain
\begin{equation}
d(z,g^{p_j}z) = \, 2\pi \left[ \left( \frac{p_j}{p} \right)^2 + \left(
G_j^2 - \left( \frac{p_j}{p} \right)^2 \right) |z_2|^2 \right]^{1/2}
\label{d} \; ,
\end{equation}
thus we need to determine the variation of $|z_2|$ inside the sphere of radius
$\chi_{obs}$ and centered at $z$.

Let an observer be located at $z \in \mathbb{S}^3$, and let $w$ be another
point at a distance $\chi_{obs} \ll 1$, then we have $\chi_{obs} > |\eta|$,
where $\eta = z - w$. Denoting by $\delta_1 = ||z_1| - |w_1||$, and
correspondingly for $\delta_2$, up to the first order in $\chi_{obs}$, one has
$|z_1| \delta_1 + |z_2| \delta_2 = 0$, which then gives $\chi_{obs}^2 >
|\eta|^2 \geq \delta_1^2 + \delta_2^2 = \left( \frac{\delta_2}{\left\vert
z_1\right\vert} \right)^2$, where the equality is achieved whenever $z_1$ is
parallel to $w_1$, and $z_2$ is parallel to $w_2$. Thus one has
\begin{equation}
\delta_2 < |z_1| \chi_{obs} \; . \label{chimax}
\end{equation}

Now in order to analyze the behavior of the isometry $g^{p_j}$, we have to
consider two different cases: (I) when the observer is not close to an equator
of minimum injectivity radius, and (II) when she is. We consider each case
separately.

\bigskip
\noindent\textbf{Case I.} $|z_1|$ and $|z_2| \gg \chi_{obs}$.
\newline

The maximum and minimum lengths of the observable geodesics associated to
$g^{p_j}$ are
\begin{equation}
d_{ext} = d_0 \pm (2\pi)^{2} \left\vert G_j^2 - \left(\frac{p_j}{p}\right)^2
\right\vert \frac{|z_2||z_1|}{d_0} \, \chi_{obs} \; , \label{dextreme}
\end{equation}
where the plus and the minus signs correspond to $d_{\max}$ and $d_{\min}$
respectively. We then have
\begin{equation}
\frac{\Delta d}{d_0} \leq 2 \, \frac{1}{|z_2||z_1|} \, \chi_{obs} \ll 1 \; .
\end{equation}
This is a bound on the maximum variation of the lengths of the smallest
geodesics in the observable universe, which means that $g^{p_j}$ behaves
CT--like. In the special case $q<\sqrt{p}$, we have seen that $p_j=1$, thus
eqs.~(\ref{dextreme}) yield
\begin{equation}
\frac{\Delta d}{d_0} \leq 4r_{inj}^2 \left( q^2 - 1 \right)
\frac{|z_2||z_1|}{d_0^2} \, \chi_{obs} \; .
\end{equation}
For $q=1$ the global homogeneity is manifest from this expression.

\bigskip \noindent\textbf{Case II.} $|z_1|$ or $|z_2| \sim
\chi_{obs}$.
\newline

If $|z_1|$ or $|z_2|~\sim1$, then as we have seen the shortest geodesics will
be generated by either $g_{(p,q)}$ or $g^{p_{k-1}}$ respectively. Both cases
are very similar. Let $z_l$ be $|z_1| $ or $|z_2|$ respectively, we then have
\begin{eqnarray*}
d_0^2 & \simeq & 4r_{inj}^2 + 2[1-\cos(2\pi\mu)] z_l^2 \; , \\
d_{ext}^2 & \simeq & 4r_{inj}^2 + 2[1 - \cos(2\pi\mu)] (z_l \pm \chi_{obs})^2
\; ,
\end{eqnarray*}
where $\mu$ is respectively $\frac{q}{p}$ or $\frac{p_{k-1}}{p}$. In these
cases the isometries $g_{(p,q)}$ and $g^{p_{k-1}}$ do not look like
translations, unless $\mu \ll 1$. Importantly, however, from an observational
point of view, the set of observers which detect such non--CT--like isometries
is small. This can be estimated by calculating the ratio of the volumes of
the $3$-sphere, $V_{M}$, and the region where $|z_l|\sim\chi_{obs}$, $V_R$, to
give
\[
\frac{V_R}{V_M} \sim \frac{3}{2} \, \chi_{obs}^2 \; ,
\]
which clearly is very small. In this way, the likelihood of an observer
detecting a non--CT--like isometry is very small.

To summarize, we have shown that subject to condition (\ref{hip}), the
detectable isometries of lens spaces behave CT--like for generic observers.
Although our results were obtained for lens spaces, they are far more
general and apply to generic spherical manifolds. This is because typically
only the largest cyclic subgroup of the covering group is detectable, resulting
in the universe `looking like a lens space', no matter what its true topology
may be~\cite{weeks}.

We now briefly discuss the consequences of our results. Recent high
resolution CMB observations are making it possible, for the first
time, to seriously look for the possible signatures of a nontrivial
cosmic topology~\cite{WMAP, Oliveira-Costa-etal2003}. They have also
produced strong support for the central predictions of inflationary
cosmology, among them the near--flatness of the universe. Together,
these provided the motivation for our study here.

Most detection methods rely on pattern repetition, and in this context
detectable isometries are those which generate closed geodesics shorter than
the maximum survey depth $\chi_{obs}$. So the subset of detectable isometries
depend crucially on both $\chi_{obs}$ and the position of the observer. We have
found that in the limit $\chi_{obs}\ll1$, the subset of such isometries in
hyperbolic and generic spherical spaces turn out to be CT--like for a typical
observer. This has the important observational consequence that a detectable
manifold is generically locally indistinguishable from cylindrical
$\mathbb{R}^2 \times \mathbb{S}^1$ (or more rarely toroidal $\mathbb{R} \times
\mathbb{T}^2$) manifolds, irrespective of its global shape. We also note that
importantly the
topological signatures expected from our results are compatible with recent
indications coming from the analyses of high resolution CMB data, such as the
alignment of the quadrupole and octopole moments of CMB anisotropies
~\cite{Tegmark-etal2003,Oliveira-Costa-etal1996}, as well as the surprisingly%
low amplitude of the CMB quadrupole~\cite{WMAP,Tegmark-etal2003}.

Given the infinite number of possible candidate manifolds any realistic search
strategy must rely on a theoretical framework which can radically restrict the
expected possibilities. Our results provide precisely such a framework by
severely restricting the detectable isometries, and thereby confining the
expected topological signatures as well as any parameter space which realistic
search strategies, such as the `circles in the sky' method, need to
concentrate on. This is particularly important in the inflationary limit, where
the order of any detectable cyclic subgroup as well as the number of candidate
isometries which generate small geodesics are extremely large. We have also
obtained bounds on $r_{max}$ which can in principle observationally rule out a
large quantity of spherical manifolds.

An important geometrical consequence of the bound $\chi_{obs} \ll 1$ is that
curvature effects are undetectable in the observable universe, resulting in
a geometrical degeneracy in this limit. Our results show that a parallel
degeneracy exists with respect to the topological structure of the universe,
in the sense that it is also impossible to distinguish topologically the
universe as being hyperbolic, flat or spherical --- in the
inflationary limit the precise nature of cosmic topology becomes
undecidable.

Our results regarding the minimal lengths of geodesics are quite general and
apply to any cyclic subgroup. The extent to which detectable isometries are
CT--like is on the other hand dependent on conditions imposed on $\chi_{obs}$.
Although we have used the condition (\ref{hip}) as synonymous to the
inflationary limit, this limit is sufficient but not necessary for (\ref{hip})
to hold. Even for far less restrictive conditions (such as observational
bounds), these results still provide useful constraints on the nature of
detectable isometries which are not necessarily CT--like. We note that the
question of likelihood of detectability does not concern us here as we are
dealing with the subset of topologies that are detectable.

Finally, our results could in principle provide a topological test for
inflation, in the sense that if observations were to detect isometries with
significant deviations from CT--likeness, they would set bounds on the amount
of inflation the primordial universe would have undergone.

A more detailed account of the results reported in this letter, as well as further
investigations of some of the questions raised here are in progress and will be
presented elsewhere.

{\bf Acknowledgments:}
We thank CNPq, FAPERJ  and FAPESP  for the grants under which this work
was carried out. We are also grateful to Jeff Weeks for his very kind
and useful correspondence.


\begin{thebibliography}{99}

\bibitem{WMAP}
D. N. Spergel \textit{et al.\/}, \texttt{astro-ph/0302209};\\
C.L. Bennet \textit{et al.\/}, \texttt{astro-ph/0302207}.

\bibitem{Oliveira-Costa-etal2003} A. de Oliveira-Costa, M. Tegmark,
M. Zaldarriaga, and A. Hamilton,  \texttt{astro-ph/0307282}.

\bibitem{Tegmark-etal2003} M. Tegmark, A. de Oliveira-Costa, and
A. Hamilton, \texttt{astro-ph/0302496}.

\bibitem{Oliveira-Costa-etal1996} A. de Oliveira-Costa, S.F. Smoot,
A.A. Starobinsky,  Astrophys.\ J. {\bf 468}, 457 (1996).

\bibitem{grt} G.I. Gomero, M.J. Rebou\c{c}as, and R. Tavakol,
Class.\ Quantum Grav.\, \textbf{18}, 4461 (2001);
Class.\ Quantum Grav.\, \textbf{18}, L145 (2001);
Int.\ J.\ Mod.\ Phys.\ \textbf{A17}, 4261 (2002); \\
B. Mota, M.J. Rebou\c{c}as, and R. Tavakol, \texttt{gr-qc/0308063},
Class. Quantum Grav.\, in press (2003).

\bibitem{weeks} J. Weeks, R. Lehoucq, and J-P. Uzan,
Class.\ Quantum Grav.\, \textbf{20}, 1529 (2003).

%\bibitem{Levin-etal-1997} J. Levin, J.D. Barrow, E.F. Bunn, and
%J. Silk, Phys.\ Rev.\ Lett.\ \textbf{79}, 974 (1997).

\bibitem{Sphere}  G.F.R. Ellis, Gen.\ Rel.\ Grav.\ \textbf{2}, 7
(1971); \\ E. Gausmann, \textit{et al.}, Class.\ Quantum
Grav.\ {\bf 18}, 5155 (2001).

\bibitem{Weeks} J. Weeks, private communication.

\bibitem{NT} A.Ya. Khinchin, \emph{Continued fractions},
University of Chicago Press, Michigan, (1964).


\end{thebibliography}
\end{document}